\documentclass[aps,twocolumn,prex,floatfix,longbibliography,superscriptaddress,pra]{revtex4-2}
\pdfoutput=1
\usepackage{float}
\usepackage[unicode]{hyperref}
\usepackage{physics}
\usepackage[export]{adjustbox}
\usepackage{enumitem}
\hypersetup{
    unicode=true, % non-Latin characters in Acrobat?s bookmarks
    plainpages=false,
    colorlinks=true,% false: boxed links; true: colored links
    linkcolor=blue,% color of internal links
    citecolor=blue,% color of links to bibliography
    filecolor=blue,% color of file links
    urlcolor=blue% color of external links
}
\usepackage{xcolor}
\usepackage{amsmath}
\usepackage[utf8]{inputenc} % allow utf-8 input
\usepackage[T1]{fontenc}    % use 8-bit T1 fonts
\usepackage{url}            % simple URL typesetting
\usepackage{booktabs}       % professional-quality tables
\usepackage{amsfonts}       % blackboard math symbols
\usepackage{nicefrac}       % compact symbols for 1/2, etc.
\usepackage{microtype}      % microtypography
\usepackage{lipsum}
\usepackage{graphicx}
\graphicspath{{media/}}     % organize your images and other figures under media/ folder
\usepackage{braket}
\usepackage[normalem]{ulem} % Include the ulem package

\begin{document}

\title{Quantum thermal machine regimes in the transverse-field Ising model}

\author{Vishnu Muraleedharan Sajitha}
\email{v.muraleedharansajitha@uq.edu.au}
\affiliation{University of Queensland -- IIT Delhi Academy of Research (UQIDAR), Hauz Khas, New Delhi 110016, India}
\affiliation{ARC Centre of Excellence for Engineered Quantum Systems, School of Mathematics and Physics, University of Queensland, St Lucia, Queensland 4072, Australia}
\affiliation{Department of Physics,
Indian Institute of Technology, Delhi, New Delhi 110016, India}
%\affiliation{The University of Queensland, Indian Institute of Technology Delhi Academy of Research (UQIDAR), New Delhi 110016, India}

\author{Bodhaditya Santra}
\affiliation{Department of Physics,
Indian Institute of Technology, Delhi, New Delhi 110016, India}

\author{Matthew J. Davis}
\email{mdavis@uq.edu.au}
\affiliation{ARC Centre of Excellence for Engineered Quantum Systems, School of Mathematics and Physics, University of Queensland, St Lucia, Queensland 4072, Australia}
\author{L. A. Williamson}
\email{lewis.williamson@uq.edu.au}
\affiliation{ARC Centre of Excellence for Engineered Quantum Systems, School of Mathematics and Physics, University of Queensland, St Lucia, Queensland 4072, Australia}

\date{\today}

\begin{abstract}
We identify and interpret the possible quantum thermal machine regimes with a transverse-field Ising model as the working substance. In general, understanding the emergence of such regimes in a many-body quantum system is challenging due to the dependence on the many energy levels in the system. By considering infinitesimal work strokes, we can understand the operation from equilibrium properties of the system. We find that infinitesimal work strokes enable both heat engine and accelerator operation, with the output and boundaries of operation described by macroscopic properties of the system, in particular the net transverse magnetization. At low temperatures, the regimes of operation and performance can be understood from the behavior of low-energy excitations in the system, while at high temperatures an expansion of the free energy in powers of inverse temperature describes the operation. The understanding generalizes to larger work strokes when the temperature difference between the hot and cold reservoirs is large. For hot and cold reservoirs close in temperature, a sufficiently large work stroke can enable refrigerator and heater regimes. Our results and method of analysis will prove useful in understanding the possible regimes of operation of quantum many-body thermal machines more generally.
\end{abstract}

\maketitle

% keywords can be removed

\section{Introduction}

A thermal machine, such as an engine or a refrigerator, consists of a working substance that utilises the flow of heat to achieve a useful task. Quantum thermal machines incorporate quantum effects in the working substance or reservoirs, providing possible performance advantages and insights into thermodynamics at the quantum scale~\cite{millen2016}. While this field has a long history in single-particle or non-interacting systems~\cite{scovil1959,10.1063/1.461951,PhysRevLett.123.240601,niedenzu2018}, recent interest has been directed toward interacting many-body quantum systems~\cite{mukherjee2021,CANGEMI20241}. Entanglement~\cite{dillenschneider2009,Abah_2014,williamson2024}, interactions~\cite{PhysRevLett.120.100601,chen2019,carollo2020,fogarty2020,boubakour2023,PhysRevB.109.024310,estrada2024,watson2024quantummanybodythermalmachines,nautiyal2024,nautiyal2025} and many-body localization~\cite{halpern2019} have been shown to enable or enhance thermodynamic tasks compared to the comparative non-interacting system.
%coherence~\cite{uzdin2015,klatzow2019,uzdin2016,tajima2021,scully2003,hardal2015,hammam2022}

As for a classical working substance, a quantum working substance may support different regimes of operation depending on the magnitude and duration of the work stroke and the temperature of the reservoirs~\cite{PhysRevB.109.224309}. However, in quantum systems, the work stroke depends on the underlying protocol, such as the two-point projective measurement scheme~\cite{PhysRevE.75.050102}, where the control parameter in the Hamiltonian is changed from an initial value to a final value. The regimes of operation for a quantum working substance  will depend on how the energies of the eigenstate change during the work stroke. Interacting quantum systems generally have a vast number of irregularly spaced energy levels. Hence, isolated work steps, even if adiabatic, can result in deviation from a thermal state due to energy levels that generally move incommensurately~\cite{quan2007,PhysRevLett.113.260601}. Therefore, understanding the regimes of operation from simple physical properties of the system is a challenging task even under adiabatic operation.
 
Arrays of interacting quantum spins are an ideal system to explore quantum many-body physics due to their rich physics and high degree of experimental control~\cite{PhysRevLett.92.207901,britton2012,bohnet2016,zhang2017,RevModPhys.93.025001,Labuhn2016,browaeys2020}, %The transverse-field Ising model~\cite{PFEUTY197079}, in particular, has been widely studied both theoretically and experimentally,
with realizations involving hundreds of spins using trapped ions~\cite{britton2012,RevModPhys.93.025001} and Rydberg atoms~\cite{Labuhn2016,browaeys2020}.
%These experimental platforms hold the potential to realize quantum many-body thermal machines.
Recently, the operation of this system as a working substance for thermodynamic tasks has become a topical area of theoretical exploration \cite{PhysRevE.100.042126,piccitto2022,revathy2020,PhysRevB.109.024310,PhysRevB.109.224309}. For nearest-neighbour interactions, performance enhancement and universal behaviour have been identified close to the quantum critical point~\cite{piccitto2022,PhysRevB.109.024310,revathy2020}. Enhancements due to long-range interactions have also been identified~\cite{wang2020,Solfanelli_2023,PhysRevB.109.024310}.

%The transverse-field Ising model with hundreds of spins can be realised experimentally using either trapped ions~\cite{britton2012,RevModPhys.93.025001} or Rydberg atoms~\cite{Labuhn2016,browaeys2020}. Interactions between trapped ions are mediated by Coulomb forces, which can be modulated via optical dipole forces, and the transverse drive is an external magnetic field. Van der Waals forces mediate interactions between Rydberg atoms, and the transverse drive is a coherent laser. However, making a manybody quantum thermal machine with transverse-field Ising model as a working medium poses serious challenges in engineering and controlling the bath and measuring outputs from it. 

%Previous studies have examined different operational regimes, focusing on how factors such as the quantum critical point ~\cite{piccitto2022} and both short- and long-range interactions ~\cite{Solfanelli_2023} influence the performance of quantum thermal machines. These studies reveal that long-range interactions can significantly enhance performance by reducing nonadiabatic losses and increasing efficiency in both the heat engine and refrigerator modes. In addition, they demonstrate how criticality can be leveraged to optimize the engine performance.

In this paper we characterise the possible regimes of adiabatic operation of a thermal machine using a spin chain with nearest-neighbor interactions as the working substance. As in previous studies, work is done on or by the system by tuning a driving field transverse to the spin interactions. We present a novel analysis based on an infinitesimal work step, which ensures the system remains in thermal equilibrium. With infinitesimal work steps, only engine or accelerator regimes are permitted. We explain the regimes of operation and the magnitude of the work output from properties of low-energy excitations and from a high-temperature expansion at low and high temperatures respectively. Boundaries between the heat engine and accelerator regimes are identified and related to the behaviour of the macroscopic magnetization of the system.

Building on the understanding provided by infinitesimal work strokes, we extend our analysis to finite-size work steps. For large differences in temperature between the cold and hot bath, the regimes of operation are qualitatively similar to the infinitesimal case, with a shift in the boundary between accelerator and heat engine operation. As the difference in the temperatures of the two reservoirs becomes small, refrigerator and heater regimes can emerge, particularly close to the quantum critical point of the system.

This paper is organised as follows. In Sec.~\ref{formalism} we introduce the model, parameterise the thermodynamic cycle, and describe the possible regimes of operation. In Sec.~\ref{results} we present our results: we present and interpret the regimes of operation and performance of a thermal machine with an infinitesimal work stroke, and then extend this analysis to a finite-size work stroke. We conclude in Sec.~\ref{conclusion}.

\section{Formalism}\label{formalism}
%\label{The Otto-cycle and different regimes of operation}
\subsection{The transverse-field Ising model}
 We consider a one-dimensional working substance consisting of $N$ spin-1/2 particles with nearest-neighbour interactions. The Hamiltonian of the system is given by the transverse-field Ising model,

\begin{equation}\label{eq:H}
    \hat{H} = -g \sum_{j=1}^{N} \hat{\sigma}_j^x \hat{\sigma}_{j+1}^x - h\sum_{j=1}^N \hat{\sigma}_j^z ,
\end{equation}
where $\hat{\sigma}_j^\mu$ are Pauli operators acting on the $j^{th}$ site ($[\hat{\sigma}_j^\mu,\hat{\sigma}_j^\nu]=2i\epsilon_{\mu\nu\kappa}\hat{\sigma}_j^\kappa$), $g>0$ is the interaction strength and $h$ is the transverse field strength. We impose periodic boundary conditions $\hat{\sigma}_{N+1} = \hat{\sigma}_1$. In the thermodynamic limit ($N\rightarrow\infty$) the transverse-field Ising model has a quantum critical point at $g=|h|$ separating the ferromagnetic ($g>|h|$) and paramagnetic ($g<|h|$) ground states \cite{Sachdev_2011}. Equation~\eqref{eq:H} can be diagonalised following a Jordan-Wigner transformation from spins to free fermions. For large $N$ this gives~\cite{PFEUTY197079}
\begin{equation}\label{eq:H2}
    \hat{H}=E_0+\sum_{j=1}^N\omega(k_j)\hat{c}_j^\dagger \hat{c}_j,
\end{equation}
with $\hat{c}_j$ the fermion annihilation operators for modes $j=1,...,N$. The fermionic mode energies are $\omega(k_j)$ with $k_j=\pi(2j-1)/N$ and
\begin{equation}\label{spectra}
    \omega(k)=2\sqrt{h^2+g^2-2gh\cos k}.
\end{equation}
The ground-state energy is $E_0=-\frac{1}{2}\sum_{j=1}^N\omega(k_j)$. 

\subsection{The quantum Otto cycle}
Among the various possible thermodynamic cycles that can be used to realise thermal machines, we choose an Otto cycle due to the ease with which heat and work can be distinguished~\cite{RAlicki_1979, 10.1063/1.446862}. The quantum Otto cycle consists of four strokes as illustrated in Fig.~\ref{Otto}(a)~\cite{e19040136}. The system begins in a hot thermal state at 1. Step (1 $\rightarrow$ 2): The system is thermally isolated and the driving field $h$ is adiabatically tuned from $h_H$ to $h_C$. Work $W_{1 \rightarrow 2}$ is done by the system. Step (2 $\rightarrow$ 3): The system is coupled to the cold reservoir with $h$ fixed, exchanging heat $Q_C$ until thermal equilibrium is achieved at 3. Step (3 $\rightarrow$ 4):  The system is thermally isolated and $h$ is adiabatically tuned from $h_C$ to $h_H$. Work $W_{3 \rightarrow 4}$ is done by the system.  Step (4 $\rightarrow$ 1): The system is coupled to the hot reservoir with $h$ fixed, exchanging heat $Q_H$ until thermal equilibrium is achieved at 1.

The total work done in the cycle is 
\begin{equation}
    W =W_{1 \rightarrow 2} + W_{3 \rightarrow 4} = -(Q_C+ Q_H).
\end{equation}
The first and second laws of thermodynamics permit four possible regimes of thermal machines, depending on the signs of heat and work \cite{PhysRevB.101.054513}, see Fig.~\ref{Otto}(b):

\begin{figure}
    \centering
    \includegraphics[width=0.37\textwidth]{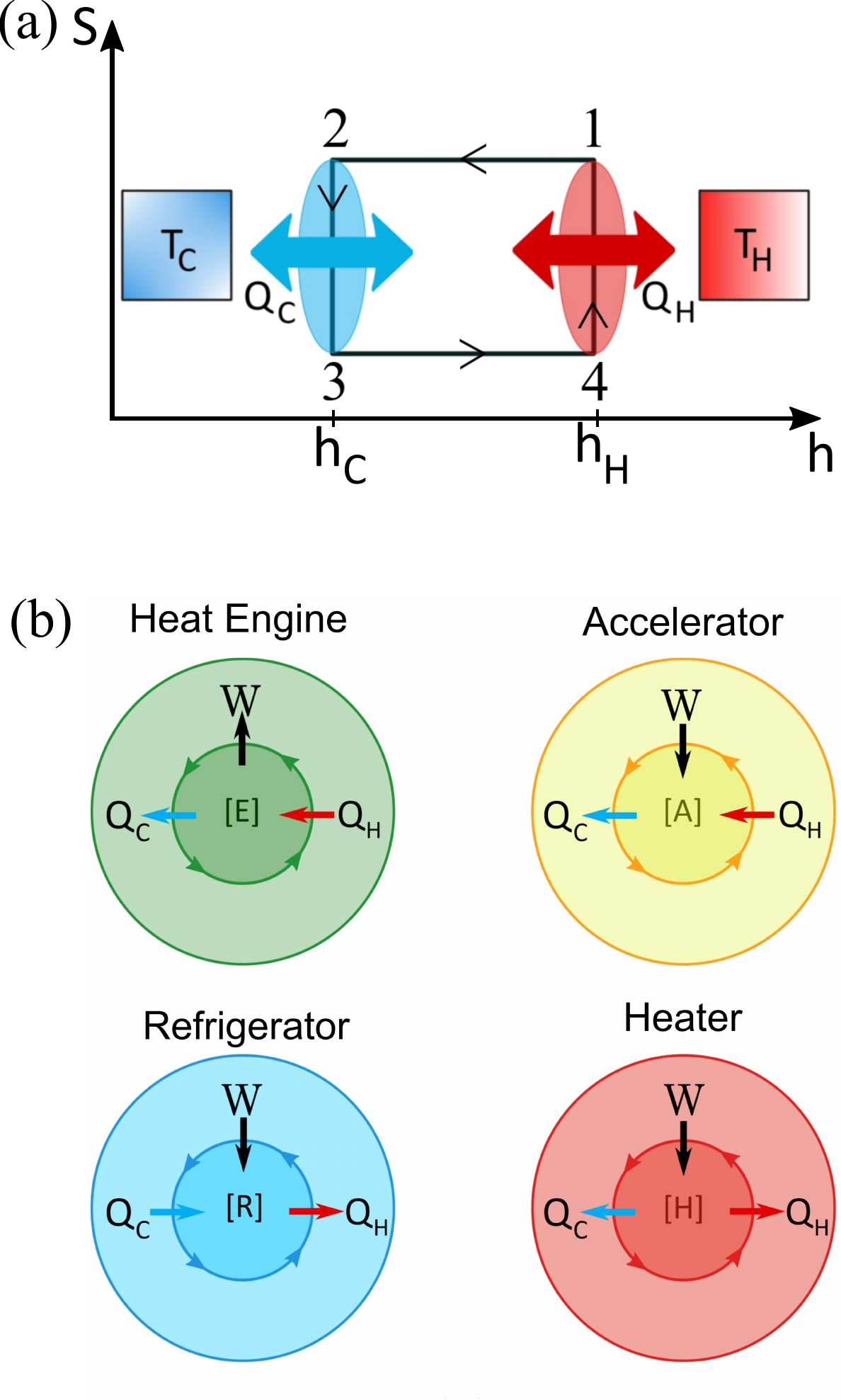}
    \caption{(a) A schematic diagram of an Otto cycle in the $S-h$ plane, where $S$ is the von-Neumann entropy. Work is exchanged by tuning the transverse field $h$ during the two unitary work strokes $1 \rightarrow 2$ and $3 \rightarrow 4$. Heat is exchanged during the two thermalization strokes $4 \rightarrow 1$ and $2 \rightarrow 3 $. (b) The four possible thermal machine operations, determined by the signs of work and heat flows.}
    \label{Otto}
\end{figure}

\begin{itemize}
    \item Engine [E]: Utilises the thermodynamic flow of heat to do work; $W < 0, Q_H >0, Q_C<0$.
    \item Accelerator [A]: Increases the thermodynamic flow of heat using work; $W > 0, Q_H >0, Q_C<0$.
    \item Refrigerator [R]: Utilises work to reverse the thermodynamic flow of heat; $W > 0, Q_H <0, Q_C>0$.  
    \item Heater [H]: Utilises work to transfer heat to both reservoirs; $W > 0, Q_H <0, Q_C<0$.
\end{itemize}
We use the convention that a negative value of heat or work is an output and a positive value an input. The thermodynamic flow of heat is a flow of heat from the hot to the cold reservoir.

In the thermodynamic limit the spacing $k_{j+1}-k_j$ between fermionic modes becomes infinitesimal, and summations $\sum_j$ over fermionic modes can be replaced by an integral $\frac{N}{\pi}\int_0^\pi dk$. The total adiabatic work output for the cycle in Fig.~\ref{Otto} is then~\cite{PhysRevB.109.224309}
\begin{equation}\label{W1}
    W=\frac{ N}{\pi}\int_0^\pi\left(\omega_C(k)-\omega_H(k)\right)  \left(f_H(k)-f_C(k)\right)dk.
\end{equation}
Here $f_{H,C}(k) \equiv [1+e^{\omega_{H,C}(k)/k_BT_{H,C}}]^{-1}$ is the Fermi-Dirac distribution, which gives the occupation of the fermionic modes in thermal equilibrium.

%We have assumed that the system completely thermalizes (to a Fermi-Dirac distribution  when coupled to the respective heat bath.

\subsection{Work and heat with an infinitesimal work step}
In general, the thermal machine regime is determined by how energy levels respond to changes in $h$ and the number of energy levels contributing (i.e.\ temperature). To simplify our initial analysis, we consider an infinitesimal work stroke $\delta h=h_H-h_C$. For $\delta h\rightarrow 0$, Eq.~\eqref{W1} simplifies to
\begin{equation}
    \begin{split}
    W=&\frac{ N}{\pi}\int_0^\pi \delta h\frac{d\omega_H(k)}{dh}\left(f_H(k)-f_C(k)\right)dk,\\
    =&\delta h\left(\left<\frac{d\hat{H}}{dh}\right>_{ T_C}-\left<\frac{d\hat{H}}{dh}\right>_{ T_H}\right).
    \end{split}
\end{equation}
%where $\left<\frac{dE}{dh}\right>_{T}$ quantifies the change in energy due to the perturbation acting on the initial thermal state of the unperturbed Hamiltonian. Similarly, the heat associated with the hot and cold reservoirs are,
Here $\langle\cdot\rangle_T$ denotes an expectation value with respect to a thermal state at temperature $T$. The quantity $\delta h\left<\frac{d\hat{H}}{dh}\right>_{T}$ is the change in system energy due to the infinitesimal work stroke $\delta h$ acting on a thermal state at temperature $T$. Similarly, the heat associated with the hot and cold reservoirs are,
\begin{equation}\label{qhc}
\begin{aligned}
Q_H &= \langle \hat{H} \rangle_{T_H} - \langle \hat{H} \rangle_{T_C} - \frac{\delta h}{k_B T_H}  \left\langle \hat{H} \frac{d\hat{H}}{dh} \right\rangle_{T_H} \\
&\quad + \delta h\left(\left<\frac{d\hat{H}}{dh}\right>_{ T_H}-\left<\frac{d\hat{H}}{dh}\right>_{ T_C}\right),\\
Q_C &= \langle \hat{H} \rangle_{T_C} - \langle \hat{H} \rangle_{T_H} + \frac{\delta h}{k_B T_H}\left\langle \hat{H} \frac{d\hat{H}}{dh} \right\rangle_{T_H}.
\end{aligned}
\end{equation}

For $\delta h\rightarrow 0$ we have $Q_H>0$ and $Q_C<0$ and hence only accelerator or engine operation is possible, depending on the sign of work. We also have the relation:
\begin{equation}
    \left<\frac{d\hat{H}}{dh}\right>_ T=\frac{dF}{dh}
\end{equation}
where $F=-k_BT\ln Z_T$ is the free energy of the system, with $Z_{ T}=\operatorname{Tr}e^{-\hat{H}/k_B T }$ the partition function at temperature $T$. Here the advantage of using an infinitesimal work stroke is clear, as it allows work and heat to be computed directly from the free energy. Using Eq.~\eqref{eq:H2} the free energy is then~\cite{PFEUTY197079}
\begin{equation}\label{freeenergy}
F =-N k_B T \left[\ln(2) + \frac{1}{\pi} \int_0^\pi \ln\left(\cosh\left(\frac{\omega(k)}{2k_B T} \right)\right)dk\right].
\end{equation}
Furthermore, using the partition function, we have
\begin{equation}\label{magnetization}
\frac{dF}{dh}= -M(T).
\end{equation}
Here $M(T)=\sum_{j=1}^N\langle\hat{\sigma}_j^z\rangle_T$ is the average transverse magnetization of the system (below we will also use the magnetization per particle $m=M/N$). Hence~\cite{PhysRevB.109.224309,PhysRevX.4.031029}
\begin{equation}
    W = \delta h ( M(T_H)-M(T_C)),
    \label{w}
\end{equation}
i.e.\ the net work exchanged is directly proportional to the difference in transverse magnetization of the system at temperatures $T_H$ and $T_C$.

Equation~\eqref{w} takes an analogous form to the work output of an ideal gas Otto cycle, which for infinitesimal volume change $\delta V$ is
\begin{equation}
    W_\text{ideal gas}=\delta V(P(T_H)-P(T_C)),
\end{equation}
with $P = -\frac{dF}{dV}$ the pressure of the system. This pressure is monotonic with temperature, resulting in operation fixed by the sign of $\delta V$. Similarly, setting $g=0$ in Eq.~\eqref{eq:H} results in operation fixed by the sign of $\delta h$. In general, for simple non-interacting systems where all energy levels vary monotonically with the control parameter, the type of operation is fixed by the sign of the change in control parameter.

%In the presence of interactions, however, non-monotonic variation in energy levels can give rise to different regimes of operation even with a fixed sign of $\delta h$.

%In both classical and quantum spin chains, the transverse magnetization may vary non-monotonically with \textcolor{red}{temperature~\cite{}}, resulting in the possibility of both engine and accelerator regimes, as we will now demonstrate.

\begin{figure*}
    \centering
    \includegraphics[width=\textwidth]{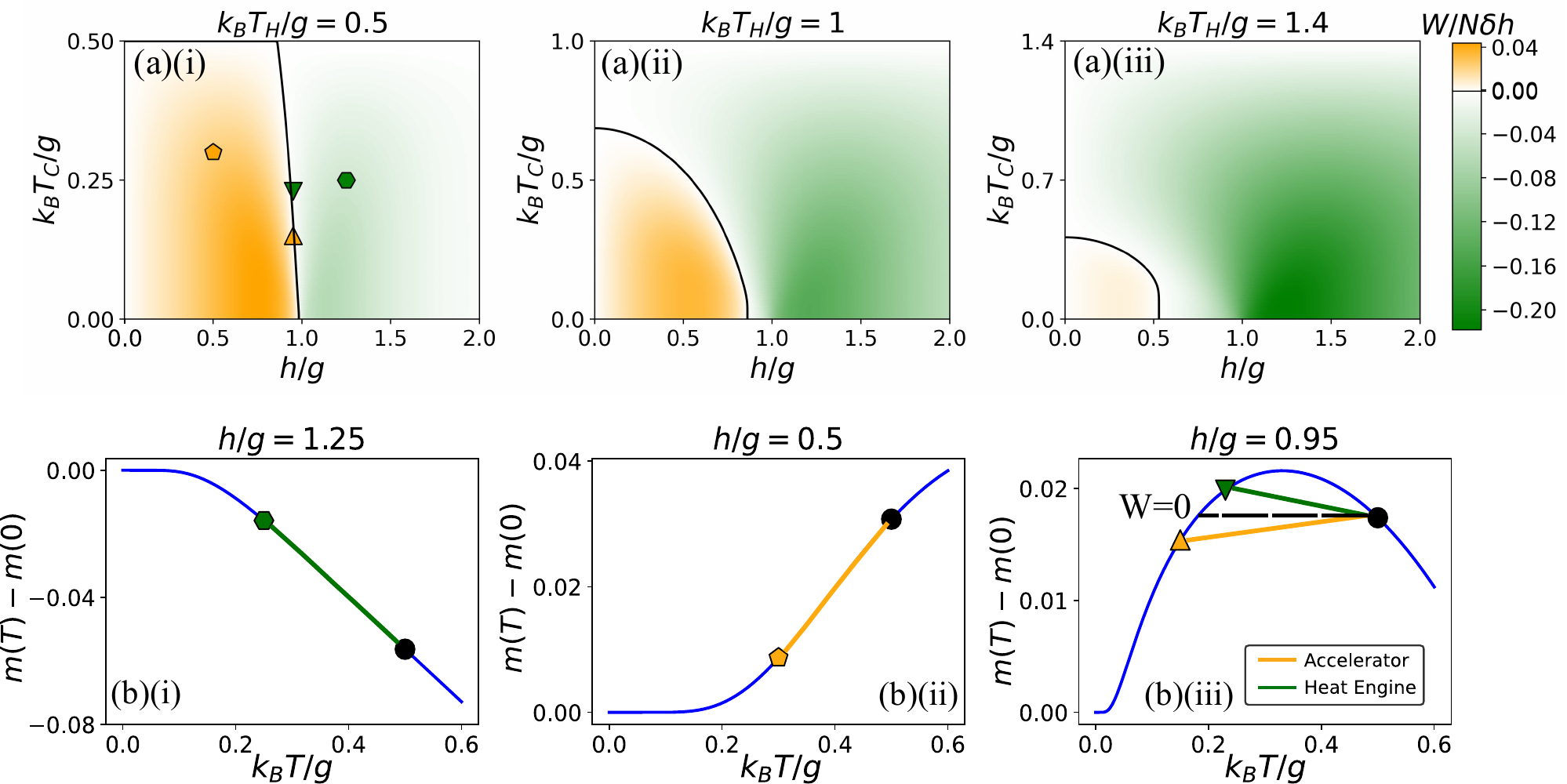}
    \caption{(a) The engine (green) and accelerator (yellow) regimes for an Otto cycle with infinitesimal work stroke $\delta h=h_H-h_C>0$ in the $ h $-$ T_C $ plane. Panels (i)–(iii) correspond to $k_BT_H/g = 0.5 $, $k_BT_H/g = 1 $ and $k_BT_H/g=1.4$, respectively. The size of work output ($W<0$) or input ($W>0$) is indicated by the colorbar. The black line represents the boundary between the engine and accelerator. (b) Transverse magnetization per particle $m(T)=M/N$ as a function of temperature for different $h/g$, with $m(0)$ the ground-state value. Black circles indicate $m(T_H)$ from (a)(i), whereas hexagon (i), pentagon (ii) and triangles (iii) indicate $m(T_C)$ corresponding to the four cycles marked in (a)(i). Engine operation occurs when $m(T_H)<m(T_C)$ whereas accelerator operation occurs when $m(T_H)>m(T_C)$. With the sign of $\delta h$ fixed, a cross-over between the two regimes requires non-monotonic behaviour in $m(T)$, as shown in (iii).}
    \label{fig:main}
\end{figure*}

\section{Results}\label{results}

\subsection{Thermal machine regimes for the transverse-field Ising model}

The work output and regimes of operation for the Otto cycle (Fig.~\ref{Otto}) with infinitesimal work step $\delta h>0$ are shown in Fig.~\ref{fig:main}(a). As already noted, with infinitesimal work step only accelerator or engine regimes are possible, with the regime of operation determined by the sign of $M(T_H)-M(T_C)$ (Eq.~\eqref{w}). When $M(T_H)<M(T_C)$ the system operates as an engine, whereas for $M(T_H)>M(T_C)$ the system operates as an accelerator. For $\delta h<0$, the sign of work is reversed and the engine and accelerator regimes are switched.

Due to the connection between work and $M(T)$, the regimes of operation can be understood from the behaviour of $M(T)$. Example behaviour of $M(T)$ are illustrated in Fig.~\ref{fig:main}(b) and Fig.~\ref{fig:magnetizationT}. The engine and accelerator regimes are separated by a boundary where $W=0$ and $M(T_H)=M(T_C)$, which requires non-monotonic behaviour in $M(T)$, see Fig.~\ref{fig:main}(b)(iii).
\subsubsection{$h>g$}
For $h>g$, the magnetization decreases monotonically as the temperature increases, see Fig.~\ref{fig:main}(b,i) and Fig.~\ref{fig:magnetizationT}. Hence $M(T_H)<M(T_C)$ and the system operates as an engine. This can be understood as follows. The ground state for $h>g$ is paramagnetic and hence will have maximum transverse magnetization. Excited levels resulting from spin-flips \cite{Sachdev_2011,mbeng2020quantum} will have small transverse magnetization, and hence as these become occupied ($T$ is increased), $M$ will decrease. More precisely, using Eqs.~\eqref{freeenergy},~\eqref{magnetization}, we have
\begin{equation}\label{dMdT}
    \frac{dM}{dT}=-\frac{4 N}{\pi k_BT^2}\int_0^\pi\left(h-g\cos k\right)\operatorname{sech}^2\left(\frac{\omega(k)}{2k_BT}\right)dk.
\end{equation}
For $h>g$ we have $dM/dT<0$ for all $T$.
\subsubsection{$h<g$ and low temperature}
\begin{figure}
    \centering
    \includegraphics[width=0.5\textwidth]{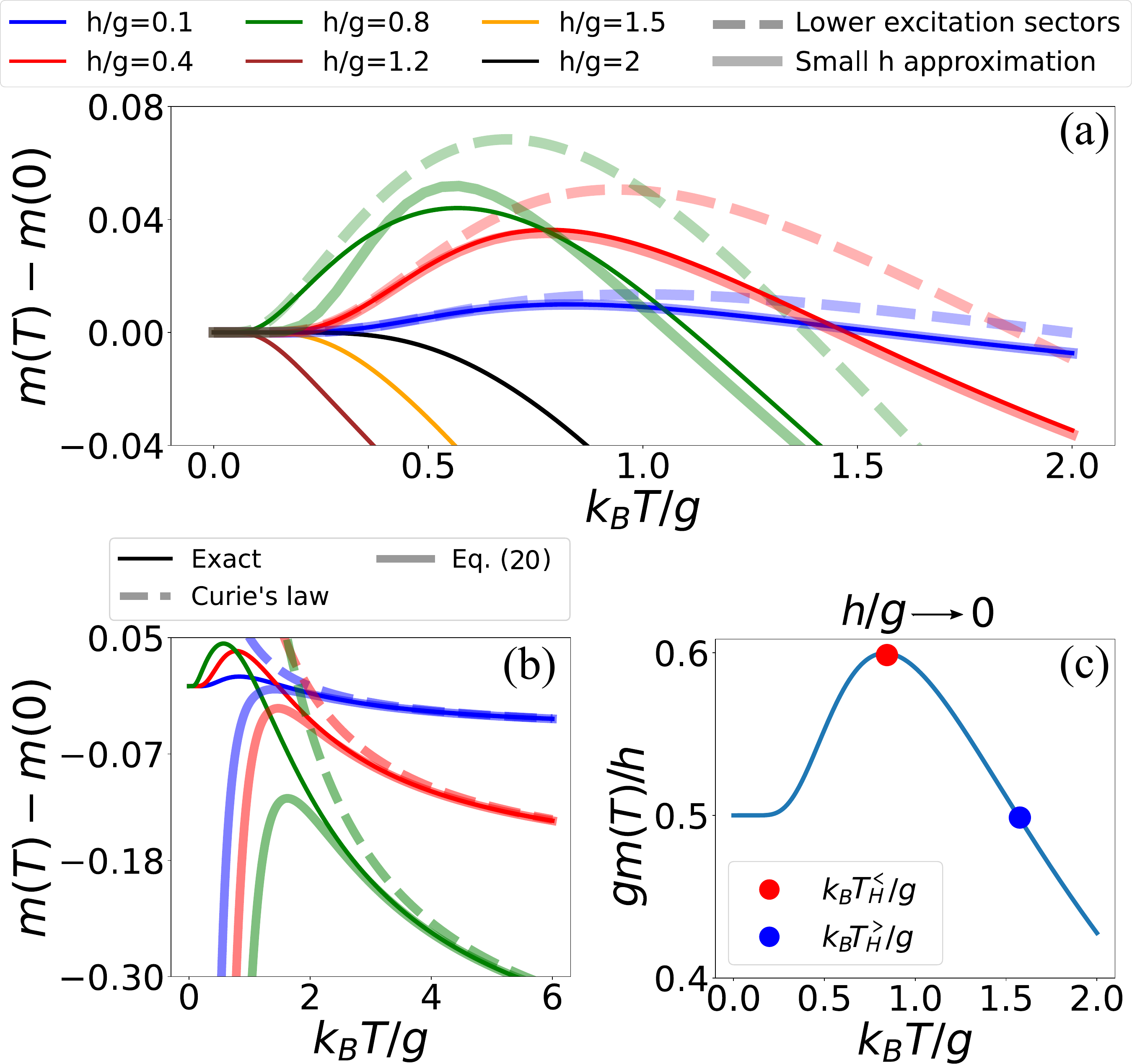}
    \caption{(a) Variation of transverse magnetization per particle with temperature showing non-monotonic behaviour that gives rise to accelerator and engine operation. Dark, thin lines are the exact result and faint thicker lines are the small-$h$ approximation Eq.~\eqref{eq:M3rd}. Dashed lines are the low-temperature approximation Eq.~\eqref{eq:MHP}. (b) The transverse magnetization decreases with increasing temperature at high temperatures, resulting in engine operation. Colors correspond to values of $h/g$ as in (a). Dark, thin lines are exact result and faint, thicker lines are the approximation Eq.~\eqref{eq:MhighT}. Dashed lines are Curie's law $m=h/k_BT$. (c) Transverse magnetization for small $h/g$, Eq.~\eqref{eq:Mlinear}, highlighting the temperatures $T_H^<$ and $T_H^>$ (see text).}
    \label{fig:magnetizationT}
\end{figure}

For $h<g$ the system can operate as an engine or an accelerator dependent on $h$ and $T$. For small $h/g$ and low temperatures we have $dM/dT>0$ and the system operates as an accelerator, see Fig.~\ref{fig:main}(b,ii) and Fig.~\ref{fig:magnetizationT}. In this regime the free energy is,  
\begin{equation}\label{eq:Fqp}
    F \approx E_0-\frac{N k_B T}{\pi} \int_0^{\pi} e^{-\frac{\mu(k)}{k_B T}}\,dk,
\end{equation}
where  $\mu(k) =  2(g-h\cos{k})$ is obtained by expanding the spectrum Eq.~\eqref{spectra} to linear order in $h/g$. The net magnetization then satisfies
\begin{equation}\label{eq:dMdTq}
    \frac{d M}{dT}\approx  \frac{2N}{\pi k_B T^2} \int_0^\pi \mu(k) e^{-\frac{\mu(k)}{k_B T}}\cos(k)\,dk>0,
\end{equation}
as anticipated.

This behaviour can be understood from domain-wall excitations above the ferromagnetic ground state. With periodic boundary conditions, these domain walls come in pairs ~\cite{Sachdev_2011} and the excitations take the form,
\begin{equation}\label{eq:states}
\ket{\psi} \approx\frac{1}{N}\sum_{\substack{j,l=1\\(j\ne l)}}^N \psi(j, l) \ket{j, l}.
\end{equation}
%\begin{equation}\label{eq:states}
%|\psi_k\rangle = \frac{1}{\sqrt{N_k}} \sum_{n=1}^{N-1} \sin(nk_k) |n\rangle,\quad k_k=\frac{k\pi}{N}, k=1,....,(N-1)
%\end{equation}
Here %$\mathcal{N}$ is the normalization factor and
%\begin{equation}
%\ket{n} = |\underbrace{-, -, \ldots, -}_{1 \rightarrow n}, +, \ldots, +\rangle \quad n = 1 \ldots N-1,
%\label{eq16}
%\end{equation}
\begin{equation}
\ket{j,l} = \ket{ +}_1\cdots\ket{+}_j\ket{-}_{j+1}\cdots\ket{-}_l\ket{+}_{l+1}\cdots\ket{+}_N
\label{eq16}
\end{equation}
is a product state with domain walls between spins $j$, $j+1$ and $l$, $l+1$, where $\ket{\pm}_j$ are the eigenstates of $\hat{\sigma}_j^x$. The $N(N-1)$ states Eq.~\eqref{eq16} are degenerate for $h=0$. The wavefunctions $\psi(j,l)$ are obtained from perturbation theory in $h$, which gives~\cite{Sachdev_2011,PhysRevB.68.212407, Rutkevich_2010,doi:10.1126/science.1180085}
%An effective Hamiltonian can be constructed with these low-lying excitations~\cite{Sachdev_2011,PhysRevB.68.212407, Rutkevich_2010}, which resembles that of a two-body hopping problem~\cite{doi:10.1126/science.1180085}. Thus,
\begin{equation}\label{psilm}
\psi(j,l) = \frac{1}{\sqrt{2}} \left( e^{i(k_1 j + k_2 l)} - e^{i(k_2 j + k_1 l)} \right),
\end{equation}
with $k_{1,2}=2\pi n_{1,2}/N$, $n_{1,2}=0,...,N-1$. The superposition in Eq.~\eqref{psilm} ensures $\psi(j, j) = 0$, as the two domain walls cannot occur at the same location. To linear order in $h$, the state Eq.~\eqref{eq:states} has energy $\mu(k_1)+\mu(k_2)=4g-2h(\cos(k_1)+\cos(k_2))$ and transverse magnetization $\braket{\psi|\sum_j\hat{\sigma}_j^z|\psi}\approx\cos k_1 + \cos k_2$. At low temperatures, increasing temperature results in increased occupation of the lower-energy domain-wall states (states with $\cos{k_1}+ \cos{k_2} > 0$), which increases the magnetization. %\textcolor{red}{Are signs correct? Increasing temp would result in higher $k$ modes occupied, which would decrease magnetization?}

\subsubsection{High temperature regime}
For high temperatures, $M(T)$ decreases monotonically with increasing temperature, see Fig.~\ref{fig:magnetizationT}(a), and engine operation is observed, see Fig.~\ref{fig:main}(a). The high-temperature behaviour of $M$ can be obtained from Eq.~\eqref{freeenergy} by expanding around $k_B T=\infty$, which gives
\begin{equation}\label{eq:MhighT}
    M=\frac{N h}{k_B T}\left[1-\frac{2g^2+h^2}{3(k_B T)^2}+O\left(\frac{1}{(k_B T)^4}\right)\right].
\end{equation}
The dominant term in Eq.~\eqref{eq:MhighT} follows Curie's law $M\propto h/k_B T$ and arises from high-temperature thermal fluctuations of the transverse spin. This results in engine operation and is present also in an ensemble of non-interacting spins. The effect of interactions on the magnetization appears at order $1/(k_B T)^3$ and requires correlations between the axial and transverse spin directions ($\partial^2 F/\partial h\partial g\ne 0$), which are suppressed at high temperatures. The result Eq.~\eqref{eq:MhighT} predicts well the exact magnetization in the monotonic regime $k_BT\gtrsim 2g$, see Fig.~\ref{fig:magnetizationT}(b).

\subsubsection{Intermediate temperature regime}
 A cross-over between the accelerator and engine regime requires non-monotonic $M$ and hence a peak where $dM/dT=0$, see Fig.~\ref{fig:main}(b)(iii). This peak is only present for $h<g$, see Fig.~\ref{fig:magnetizationT}(a), and separates the low and high temperature regimes described above. The crossover ($W=0$ line) can be determined by solving $M(T_C)=M(T_H)$. To lowest order in $h/g$ we have
 \begin{equation}\label{eq:Mlinear}
     M(T)= \frac{ Nh}{2g} \left[\tanh\left(\frac{g}{k_BT}\right)+ \frac{g}{k_B T} \text{sech}^2\left(\frac{g}{k_BT}\right)\right],
 \end{equation}
which is plotted in Fig.~\ref{fig:magnetizationT}(c). From this plot it is clear that $M(T_H)=M(T_C)$ has a non-trivial solution $T_H\ne T_C$ only for $T_H^<<T_H\le T_H^>$. Here $k_BT_H^</g\approx 0.83$ is the temperature the magnetization peaks, obtained from Eq.~\eqref{eq:Mlinear} by solving $dM/dh=0$. The higher temperature $k_BT_H^>/g=1.56$ is obtained from Eq.~\eqref{eq:Mlinear} by solving $M(T)=M(0)$. For small $h/g$ the system operates strictly as an accelerator for $T_H<T_H^<$ and as an engine for $T_H>T_H^>$.
 
 %which give the transcendental equations,
%\begin{equation}
%    \begin{aligned}
%        \frac{g}{T_H^<}\operatorname{sech}^2\frac{g}{T_H^<}+\tanh\frac{g}{T_H^<}-1=0\\
%        \frac{g}{T_H^>}\tanh\frac{g}{T_H^>}-1=0
%    \end{aligned}
%\end{equation}.
%with numerically obtained solutions as above. 

With increasing $h/g$, the $W=0$ boundary tends to move to lower $T_C$, see Fig.~\ref{fig:main}(a). For $h\rightarrow g$, the $W=0$ point occurs at $T_C,T_H\rightarrow 0$. A change in sign of $dM/dT$ requires a change in sign of $d\omega(k)/dh$ within the range of thermally accessible $k$. We have $dM/dT=0$ when
\begin{equation}
    \frac{d\omega(k)}{dh}=\frac{2(h-g\cos k)}{\sqrt{g^2+h^2-2g h\cos k}}\approx 0
\end{equation}
for some $\omega(k)\approx k_B T$. From this we can estimate that the $W=0$ line occurs when $\omega(k)|_{h=g\cos k}=\sqrt{g^2-h^2}\sim k_B T$, giving rise to a $W=0$ line that moves to lower $T_C$ as $k_BT_C\sim \sqrt{g^2-h^2}$. Note the $W=0$ boundary requires using a spectrum beyond the perturbative approximation Eq.~\eqref{eq:dMdTq}, which gives $dM/dT>0$. Replacing $\mu(k)$ by $\omega(k)$ in Eq.~\eqref{eq:Fqp} (equivalently expanding Eq.~\eqref{freeenergy} in powers of $e^{-\omega(k)/k_B T}$) gives
%\begin{equation}\label{eq:FreeLT}
%    F =-\frac{N k_B T}{\pi} \int_0^{\pi}dk e^{-\frac{\omega(k)}{k_B T}}
%\end{equation} 
%with the transverse magnetization given by,
\begin{equation}\label{eq:MHP}
   M(T) =M(0) -\frac{4N}{\pi} \int_0^{\pi} \frac{h-g\cos(k)}{\omega(k)}e^{-\frac{\omega(k)}{k_B T}}\,dk.
\end{equation}
The approximation Eq.~\eqref{eq:MHP} qualitatively predicts the peak in magnetization for $h<g$, see Fig.~\ref{fig:magnetizationT}(a).

A more accurate expression for $M$ for $h<g$ can be obtained by expanding the full expression for $M$ to third order in $h/g$. The resulting expression, however, is complicated,
\begin{equation}\label{eq:M3rd}
\begin{split}
    M=& \frac{ Nh}{2g} \left(\tanh x + x \text{sech}^2 x\right)- \frac{N h^3}{16 g^3} \Big[6x^3 \text{sech}^4 x\\&- \tanh x+x\text{sech}^2 x\left(1 - 4x^2 + 4x \tanh x\right)\Big],
\end{split}
\end{equation}
with $x=g/k_B T$. The approximation Eq.~\eqref{eq:M3rd} agrees well with the exact magnetization for $h/g\lesssim 0.5$, see Fig.~\ref{fig:magnetizationT}(a).

\begin{figure}[t]
    \centering
    \includegraphics[width=0.45\textwidth]{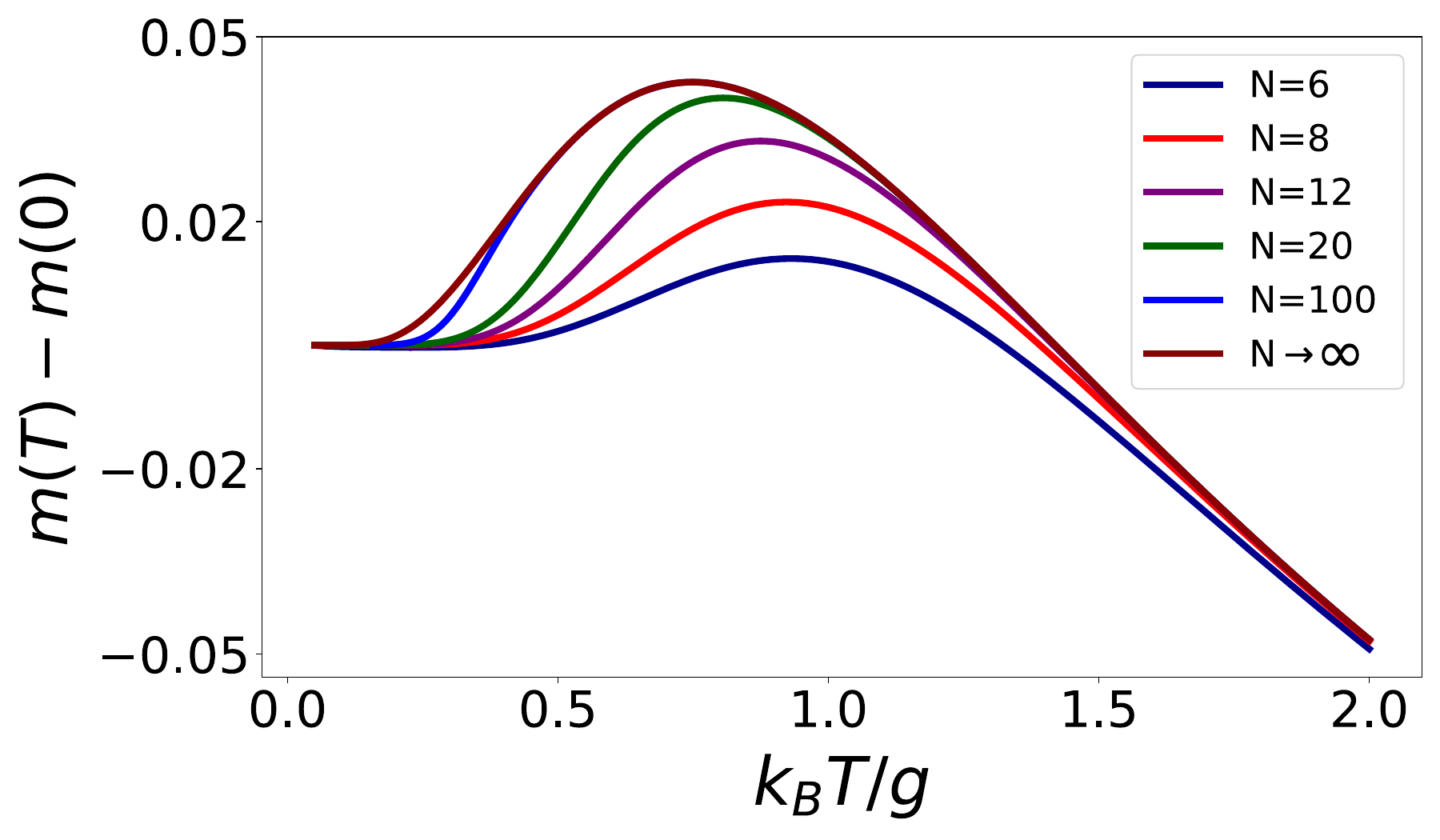}
    \caption{The transverse magnetization per particle as a function of $k_BT/g$ for finite size systems with $h/g = 0.5$, compared to the thermodynamic limit. The curves are qualitatively similar and hence so are the regimes of thermal machine operation. We find $m(T)-m(0)$ decreases with decreasing $N$ such that the temperature range for which both accelerator and engine operation is possible narrows.}
    \label{fig:magnetizationN}
\end{figure}

\subsubsection{Finite-size effects}

The free energy for the transverse field Ising model can also be calculated for a finite number of spins after accounting for parity~\cite{PhysRev.127.1508,10.21468/SciPostPhys.11.1.013}. The magnetization for systems with a finite number of spins and $h<g$ are compared with the thermodynamic limit in Fig.~\ref{fig:magnetizationN}. For $N\gtrsim 100$ the magnetization per particle agrees well with the thermodynamic limit. For smaller $N$ the magnetization is qualitatively similar to the thermodynamic limit, and results in strictly accelerator operation for sufficiently low temperatures of the hot bath and engine operation for sufficiently high temperatures [c.f.\ Fig.~\ref{fig:magnetizationT}(c)]. We find $m(T)-m(0)$ decreases with decreasing $N$; hence the temperature range separating the strictly engine and accelerator regimes narrows as the system size is reduced, see Fig.~\ref{fig:magnetizationN}. The decrease in $m(T)-m(0)$ with $N$ occurs because the sparsely spaced energy levels in finite-size systems require a higher temperature to obtain a given transverse magnetization. We observe a cross-over to monotonically decreasing $m(T)$ for $h\approx g$, as for the thermodynamic limit, with the precise cross-over sensitive to $N$. As $h/g$ increases further the effect of interactions decreases and the finite-size results for $m(T)$ converge to the thermodynamic limit even for small $N$.

%\begin{equation}
%\begin{split}
%    M &= \frac{ h N}{16 g^3} \biggl(6 g^3 h^2 \frac{1}{k_B T}^3 \text{sech}^4(\frac{g}{k_B T})- (8 g^2 + h^2) \tanh(\frac{g}{k_B T}) \\
%    &\quad + \frac{g}{k_B T} \text{sech}^2(\frac{g}{k_B T}) \bigl(h^2 - 4 g^2 (2 + h^2 \frac{1}{k_B T}^2) + 4 g h^2 \frac{1}{k_B T} \tanh(\frac{g}{k_B T})\bigr)\biggr)
%\end{split}
%\end{equation}

\begin{figure}
    \centering
    \includegraphics[width=0.5\textwidth]{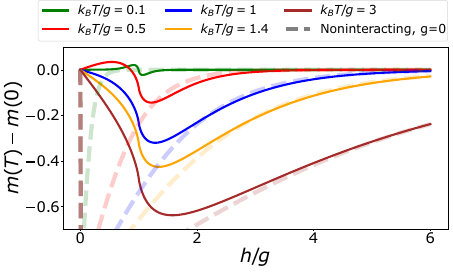}
    \caption{The transverse magnetization per particle as a function of $h/g$. We can see $m(T)-m(0)$  exhibits a minimum near $h = g$ and therefore results in a maximum for $|W| = \delta h( m(T_H)-m(T_C))$, see Eq.~\eqref{w}. The magnetization approaches the non-interacting result Eq.~\eqref{eq:Mnonint} (dashed line) at large $h/g$; in this regime $|W|$ decreases with increasing temperature.}
    \label{fig:magnetizationh}
\end{figure}

 %Using Eqs.~\eqref{freeenergy},~\eqref{magnetization} we have
%\begin{equation}\label{hstar}
 %   M(T_H)-M(0)=\frac{N}{2\pi}\int_0^\pi dk\,\frac{h-g\cos k}{\omega(k)}\left(1-\tanh\left(\frac{1}{2}\frac{1}{k_B T}\omega(k)\right)\right).
%\end{equation}
%Expanding Eq.~\eqref{hstar} around $\frac{1}{k_B T}=\infty$ and $\omega(k)$ around $h=g$ gives
%\begin{equation}
%    M(T_H)-M(0)=\int_0^\pi dk\,\frac{h^*-g\cos k}{\omega(k)}e^{\frac{1}{k_B T}\omega(k)}
%\end{equation}
%\textcolor{red}{add in more manipulation... Prefactors are wrong}
%Since..., solving $M(T_H)-M(0)=0$ for $h^*$ gives $h^*\rightarrow g$ as $T_H\rightarrow 0$ and $h^*$ decreasing with increasing $T_H$.

\subsection{Magnitude of work output}

\begin{figure*}[t]
    \centering
    \includegraphics[width=\textwidth]{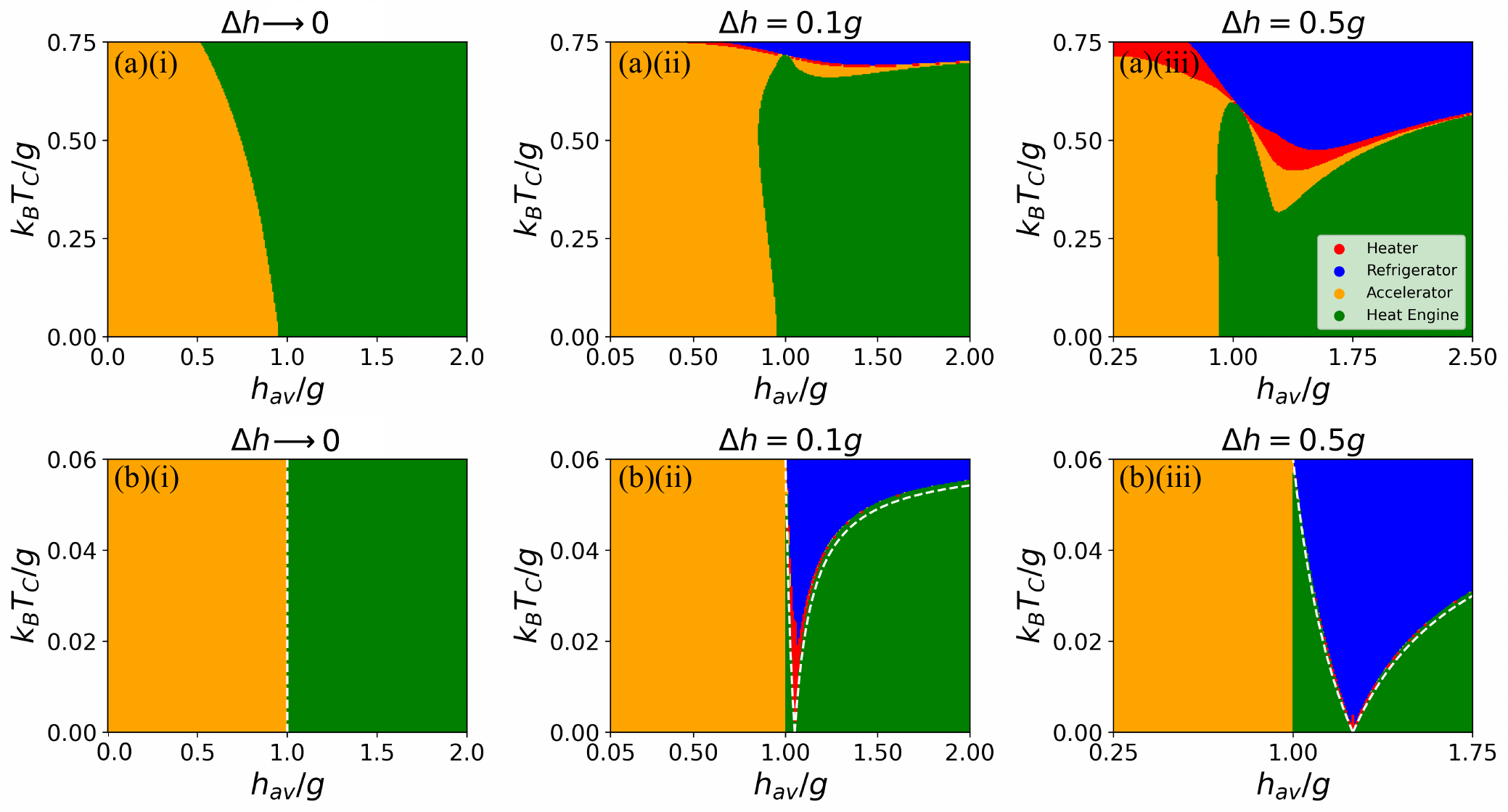}
    \caption{(a) Thermal machine regions for increasing $\Delta h=h_H-h_C$, with (a) $k_B T_H/g = 0.75$ and (b) $k_B T_H/g = 0.06$. Finite-size $\Delta h$ gives rise to heater and refrigerator regimes. White dashed lines in (b) represent the boundaries derived from the inequalities Eq.~\eqref{ineq1}.}
    \label{phasecompfinite}
\end{figure*}

We now briefly discuss the magnitude of work extracted (engine) or consumed (accelerator). In the high-temperature, low-$h$ limit, $|W|$ increases with increasing $h$, see Fig.~\ref{fig:main}. This can be seen directly from the high temperature expansion Eq.~\eqref{eq:MhighT}, which also results in $|M(T_H)-M(T_C)|$ decreasing with increasing $T$ for fixed $T_H/T_C$. In contrast, in the low-temperature, high-$h$ limit, $|W|$ decreases with increasing $h$, see Fig.~\ref{fig:main}. When $ h\gg g $, the transverse magnetization can be approximated by the magnetization of non-interacting spins, 
\begin{equation}\label{eq:Mnonint}
    M\approx N \tanh\left(\frac{h}{k_B T}\right),
\end{equation}
For low temperatures $k_B T\ll h$ this gives $M\approx N(1-2e^{-h/k_B T})$, in which case we find $|M(T_H)-M(T_C)|$ increases with increasing $T$ for fixed $T_H/T_C$. Both of these arguments also hold in the absence of interactions. %In temperature regions where $M(T)$ is monotonic $|M(T_H)-M(T_C)|$, and hence $|W|$, will clearly increase with increasing $T_H-T_C$.

For $k_B T\lesssim g$, $|M(T)-M(0)|$ is largest around $h\approx g$, see~Fig. \ref{fig:magnetizationh}. This results in a peak in $|W|$ for $h\approx g$, see Fig.~\ref{fig:main}. Such a peak has been discussed already in the low-temperature regime, in which case it can be explained in terms of a decreasing energy gap between the ground and first excited state~\cite{PhysRevB.109.024310}. At low temperatures and for $h>g$, this results in a performance that exceeds that of non-interacting spins~\cite{PhysRevB.109.024310}.

\subsection{Finite-size work strokes}
For an infinitesimal work stroke, the sign of the heat flows are fixed, restricting operation to either engine or accelerator, see Eq.~\eqref{qhc}. A finite-sized work stroke $\Delta h=h_H-h_C$ permits the thermal machine to operate as a refrigerator or a heater, see Fig.~\ref{phasecompfinite}, where we have defined $h_\mathrm{av}=(h_H+h_C)/2$.

To see how refrigeration operation may appear, it is simplest to first consider very low temperatures, in which case only the ground and first excited states have any significant thermal occupation~\cite{PhysRevB.109.024310}. The analysis then proceeds as for a single spin-1/2 system~\cite{PhysRevB.101.054513}. For low temperatures the heat flows are,
\begin{equation}
\begin{aligned}
    Q_H &\approx 2 |h_H-g| \left(e^{-|h_H-g|/k_BT_H} - e^{-|h_C-g|/k_B T_C}\right), \\
    Q_C &\approx 2 |h_C-g| \left(e^{-|h_C-g|/k_B T_C}-e^{-|h_H-g|/k_B T_H}\right)
    \label{20}
\end{aligned}
\end{equation}
with $|h-g|$ the energy of the first excited level, which is the $k=0$ fermionic mode.
From examination of Eq.~\eqref{20}, it is clear that low-temperature refrigeration ($Q_H<0$, $Q_C>0$) occurs when
\begin{equation}\label{eq:fridge}
  |h_H-g|>\frac{T_H}{T_C}|h_C-g|.
\end{equation}
For $\Delta h>0$ Eq.~\eqref{eq:fridge} is satisfied when
\begin{equation}
 g + \frac{\Delta hT_H}{T_H + T_C}< h_H < g + \frac{\Delta h T_H}{T_H-T_C}.
  \label{ineq1}
\end{equation}
The bounds of the inequality~\eqref{ineq1} accurately predict the low-temperature refrigerator boundary, see Fig.~\ref{phasecompfinite}(b). Similar inequalities can be obtained from Eq.~\eqref{eq:fridge} for $\Delta h<0$.

For higher temperatures, the analysis is qualitatively similar but complicated by the many levels present in the problem. This case has been discussed in recent works~\cite{Solfanelli_2023, piccitto2022, PhysRevB.109.224309}. Notably, refrigerator operation is more effective when the temperature difference between $T_H$ and $T_C$ is small, and the influence of the critical point results in a peak in cooling capability~\cite{Solfanelli_2023}. Additionally, it was shown that as the size of the work stroke increases, the boundary of the refrigerator region expands to smaller $T_C$~\cite{piccitto2022}. This expansion of the refrigerator region becomes more pronounced with work strokes across the critical point, as seen in Fig.~\ref{phasecompfinite}.

 All four regimes of operation appear for finite $\Delta h$ at sufficiently high temperature, see Fig.~\ref{phasecompfinite}(a). This gives rise to a ``Carnot point'' where $Q_H=Q_C=W=0$ and all four regimes intersect~\cite{PhysRevB.101.054513,PhysRevB.109.224309}. A sufficient condition for $Q_H=Q_C=W=0$ is $r(k)=\omega_H(k)/\omega_C(k)=T_C/T_H$ for all $k$~\cite{PhysRevB.109.224309}. The ratio $r$ is insensitive to $k$ when $dr(k)/dk=0$, which gives $g^2 = h_H h_C$ [using Eq.~\eqref{spectra}]. Substituting this into $r$ and setting $r=T_C/T_H$ gives values $h_C^\mathrm{cp}$ and $h_H^\mathrm{cp}$ where $Q_H=Q_C=W=0$ for given $T_C$, $T_H$ and $g$~\cite{PhysRevB.109.224309},
%\begin{equation}
%    r = \frac{h_C}{g} = \frac{g}{h_H} = \frac{T_C}{T_H}.
%    \label{CP}
%\end{equation}
\begin{equation}
    h_C^\mathrm{cp}=\frac{g T_C}{T_H},\hspace{1cm}h_H^\mathrm{cp}=\frac{gT_H}{T_C}.
    \label{CP}
\end{equation}

\begin{figure}
    \centering
    \includegraphics[width=\linewidth]{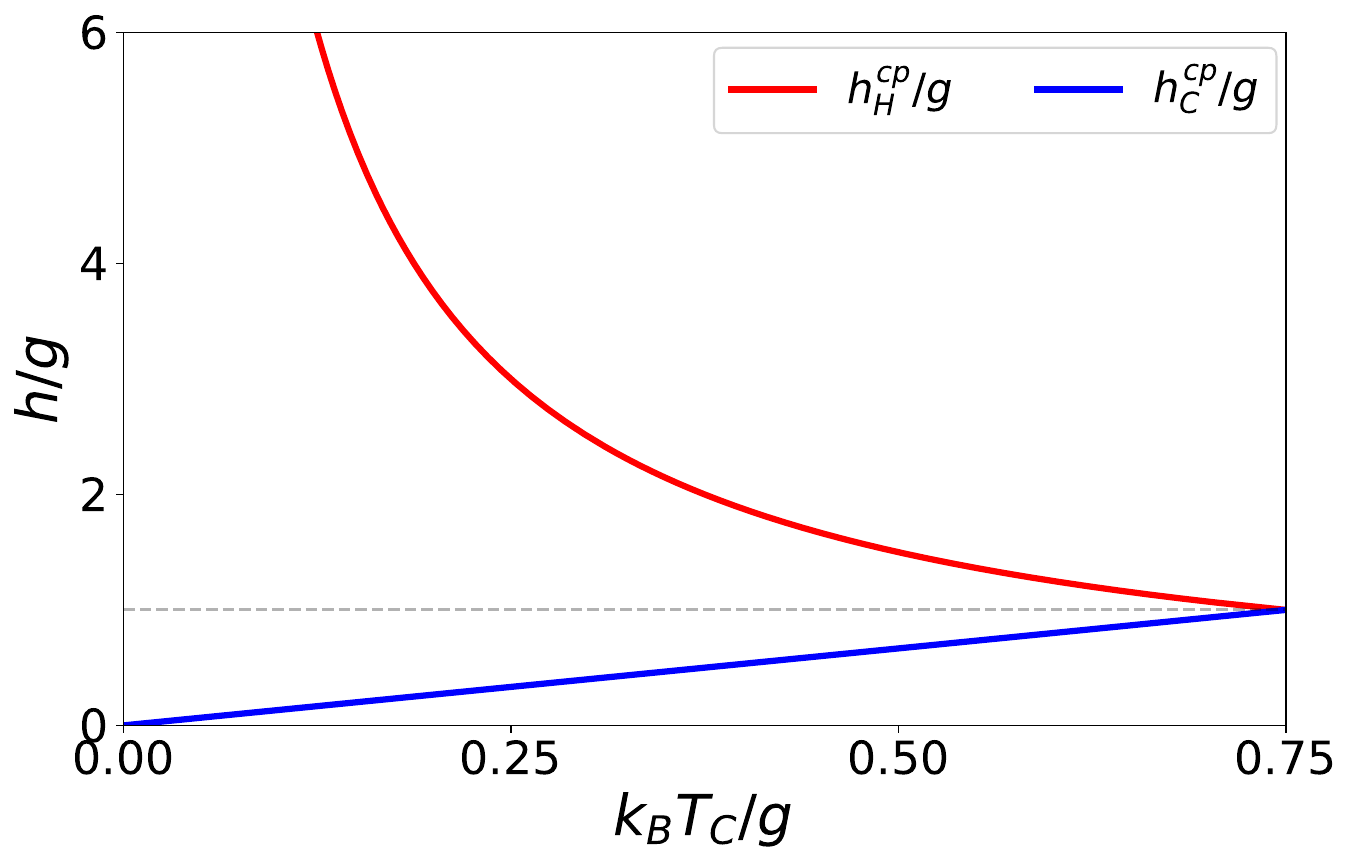}
    \caption{Values of the driving  $h_H^\mathrm{cp}$ and $h_C^\mathrm{cp}$ for a cycle with $Q_H=Q_C=W=0$ [Eq.~\eqref{CP}], which gives rise to Carnot point where all four regimes of operation intersect, see Fig.~\ref{phasecompfinite}(a)(ii) and (iii). We have $h_C^\mathrm{cp}<g<h_H^\mathrm{cp}$ and hence the work stroke necessarily crosses $h=g$ (horizontal dashed line). The temperature of the hot bath is fixed at $k_BT_H/g = 0.75$. We have also calculated $h_C^\mathrm{cp}$ and $h_H^\mathrm{cp}$ numerically using Eq.~\eqref{W1} and similar expressions for heat, which gives results indistinguishable from Eq.~\eqref{CP}.}
    \label{fig:CarnotPoint}
\end{figure}

The field strengths $h_H^\mathrm{cp}$ and $h_C^\mathrm{cp}$ are shown in Fig.~\ref{fig:CarnotPoint}. The Carnot point appears when $\Delta h$ crosses $g$, since $h_C^\mathrm{cp}<g<h_H^\mathrm{cp}$. This is reasonable because the sign of $\frac{d\omega}{dh}$ is most sensitive to $h$ near $g$, so small variations in $h_\mathrm{av}$ can lead to changes in the sign of $W, Q_H$ and $Q_C$. For small $\Delta h$, the Carnot point will appear when $T_H\approx T_C$, as then only a small $\Delta h$ is needed to change the flow of heat, see Fig.~\ref{phasecompfinite}(a)(ii). As $\Delta h$ increases ($h_H\gg g$), the accelerator and heater regimes shrink, and the system behavior converges to that of a noninteracting spin chain, where only the engine and refrigerator regimes remain. 

When $k_BT_H-k_BT_C \gg |\Delta h|$ the sign of the heat flows are fixed and only accelerator or engine operation is possible. Furthermore, the $W = 0$ boundary separating these two regimes is unaffected by $\Delta h$ at low temperatures, see Fig.~\ref{phasecompfinite}. Focusing on just the ground and first excited state, as in Eq.~\eqref{20}, zero work output will occur when $|h_C-g|=|h_H-g|$, in which case the work stroke causes no net change in the energy of the first excited state. This gives $h_\mathrm{av}=g$ irrespective of $\Delta h$.
%For $r_{k}%(h_H,h_C)$ to be constant we need

%\begin{equation}
%    \pdv{r^2_{k}(h_H,h_C)}{k} = \frac{2 g \sin(k) (h_H - h_C) (g^2 - h_C h_H)}{(g^2+ h_C^2 - 2 g h_C \cos(k)  )^2} = 0.
%\end{equation}

\section{Conclusion}\label{conclusion}

We have presented the quantum thermal machine regimes for the transverse-field Ising model with an infinitesimal work stroke. In this scenario the heat flows are fixed by the temperatures of the hot and cold reservoirs.  This results in either the heat engine or accelerator regime, dependent on the difference in equilibrium transverse magnetization at the two temperatures. We have identified the physical mechanisms behind the regimes of operation, connecting the low-temperature operation to the behavior of low-energy excitations and the high-temperature operation to an approximate equation of state for the system. This qualitative understanding also explains the regimes of operation for a finite-size work stroke when the difference in hot and cold reservoir temperatures are sufficiently large relative to the work stroke. Otherwise, refrigerator and heater regimes can emerge. Although most of our analysis here has been done in the thermodynamic limit, we have also shown that similar results hold in finite-size systems, as would be realised experimentally.%At low temperatures the presence of these additional regimes can be explained in terms of quasiparticles, providing an analytic estimate for the boundaries between the different regimes of operation.

The realization of the transverse-field Ising model with either trapped ions or Rydberg atoms offers the potential to experimentally implement a many-body quantum thermal machine. Interactions between trapped ions are mediated by Coulomb forces, which can be modulated via optical dipole forces, and the transverse drive is an external magnetic field. Van der Waals forces mediate interactions between Rydberg atoms, and the transverse drive is a coherent laser. In both setups the work step can easily be implemented by varying the intensity of the transverse drive. Controlled heating and cooling of the system poses a challenge, but may be possible by applying external noise or light beams~\cite{Robnagel,PhysRevLett.119.050602}. The work done by the spins changes the power of the transverse drive, but the change is too small to be experimentally detectable~\footnote{To see this, first note that adiabatic engine operation requires a cycle time $\gtrsim h^{-1}$ and therefore the change in power $P$ of the drive is $|\delta P|\lesssim h^2$. For a system driven by a magnetic field $B$ we have $h=\mu B$, with $\mu$ the magnetic dipole moment of the spins, while for a system driven by an electric field $E$ we have $h=d E$, with $d$ the electric dipole moment. Estimating $\mu\sim \mu_B$ and $d\sim e a_B$, with $\mu_B$ the Bohr magneton, $a_B$ the Bohr radius and $e$ the electron charge, gives $|\delta P|/P\lesssim N\alpha^3 a_B^2/A$ for a magnetic drive and $|\delta P|/P\lesssim N\alpha a_B^2/A$ for an electrical drive, with $\alpha=e^2/(4\pi\epsilon_0\hbar c)$ the fine structure constant, $A$ being the area of the drive and $P=(1/2)\epsilon_0 cE^2A$. For an electric drive with a micron scale beam waist we have $|\delta P|/P\lesssim 10^{-11}N$; for a magnetic drive the output is even smaller.}. The work done by the spins could instead by inferred from measuring the change in energy of the working substance itself~\cite{Onishchenko2024} using site-resolved imaging~ \cite{Labuhn2016,Islam2011}.

Our methodology and analysis will be useful in exploring other complex many-body quantum thermal machines, such as spin chains with long-range interactions and interacting Bose gases~\cite{CANGEMI20241}. Diabatic work steps will likely modify the regimes of operation~\cite{PhysRevB.101.054513}. Considering infinitesimal work strokes would make diabatic operation amenable to a perturbative analysis~\cite{PhysRevResearch.2.023377} and allow counterdiabatic protocols to be incorporated~\cite{PhysRevLett.111.100502}, providing an interesting avenue for future research.\\

\begin{acknowledgments}

This research was supported by The University of Queensland--IITD Academy of Research (UQIDAR), the Australian Research Council Centre of Excellence for Engineered Quantum Systems (CE170100009), and the Australian federal government Department of Industry, Science, and Resources via the Australia-India Strategic Research Fund (AIRXIV000025). We also acknowledge the support from the Indian Institute of Technology Delhi and SERB-DST, India.
\end{acknowledgments}

\end{document}